**Running head:** Evolving Decision and Affect in LLMs

# Developmental trajectories of decision making and affective dynamics in large language models

Zhihao Wang[1], Yiyang Liu[2,1] [*], Ting Wang[3,*], Zhiyuan Liu[1,*]

[1]Center for Neurocognition and Social Behavior, Institute of Artificial Intelligence, Shenzhen University of Advanced Technology, Shenzhen, 518055, China
[2]State Key Laboratory for Novel Software Technology, Nanjing University, Nanjing, China
[3]Institute for brain research and rehabilitation, South China Normal University, Guangzhou, China

[*]Corresponding authors:

Yiyang Liu
State Key Laboratory for Novel Software Technology, Nanjing University, Nanjing, 200023, PR China
Email: liuyiyang@suat-sz.edu.cn

Ting Wang, Ph.D.
Institute for brain research and rehabilitation, South China Normal University, Guangzhou, 510631, PR China
Email: t.wang@m.scnu.edu.cn;

Zhiyuan Liu, Ph.D.
Center for Neurocognition and Social Behavior, Institute of Artificial Intelligence, Shenzhen University of Advanced Technology, Shenzhen, 518055, PR China
Email: liuzhiyuan@suat-sz.edu.cn.

Number of pages: 23
Number of figures: 3, Table: 0
Number of words for abstract: 148, introduction: 648, discussion: 1187[1]

---

[1]The authors declare no competing financial interests.

**Abstract**

Large language models (LLMs) are increasingly used in medicine and clinical workflows, yet we know little about their decision and affective profiles. Taking a historically informed outlook on the future, we treated successive OpenAI models as an evolving lineage and compared them with humans in a gambling task with repeated happiness ratings. Computational analyses showed that some aspects became more human-like: newer models took more risks and displayed more human-like patterns of Pavlovian approach and avoidance. At the same time, distinctly non-human signatures emerged: loss aversion dropped below neutral levels, choices became more deterministic than in humans, affective decay increased across versions and exceeded human levels, and baseline mood remained chronically higher than in humans. These "developmental" trajectories reveal an emerging psychology of machines and have direct implications for AI ethics and for thinking about how LLMs might be integrated into clinical decision support and other high-stakes domains.

**Introduction**

Large language models (LLMs) have rapidly become embedded in everyday life and professional practice[1], with particularly striking potential in medicine and health care[2]. Beyond automating administrative tasks, LLMs are increasingly integrated into clinical decision support systems, where they provide diagnostic suggestions, treatment options, and patient-facing explanations[3–5]. In these contexts, LLMs are not only processing information but effectively participating in high-stakes decision processes: for example, helping to weigh conservative versus aggressive treatment options, or framing risks and benefits for patients[3]. At the same time, their conversational style – including how they express concern, reassurance, or empathy – can strongly shape patients' emotional experiences and perceived quality of care[5–7]. As LLMs are deployed at scale in such consequential settings, a natural question arises: what kind of "psychology" are we inviting into our decision processes?

Most existing evaluations of LLMs approach this question from a static perspective[8]. A growing literature has assessed logical reasoning[9–11], decision making[12,13], and even affective or "emotional" capacities[14,15] in a single model at a given time, or by qualitatively comparing two contemporaneous systems. However, modern LLMs are evolving at an unprecedented pace. Successive generations (e.g., GPT-3.5, GPT-4, GPT-4o, GPT-4.1, and beyond) differ not only in benchmark performance but also in their emergent behavioral profiles[16]. This raises a critical concern about the temporal generalizability of current findings [9,12,17,18]: conclusions drawn about "GPT-4" in 2024 may be of limited relevance by the time most users have migrated to a newer version. To address this gap, we draw on the idea of a historically informed outlook on the future and are inspired by human developmental psychology. Specifically, we adopt a developmental perspective to systematically probe how risk-related decision behavior and affective output style change across successive generations of LLMs within the same model family.

In human research, the gold standard for precisely assessing cognitive processes has moved beyond self-report questionnaires, which are susceptible to bias and social desirability effects[19,20]. Instead, the field increasingly prioritizes the computational modeling of behavior and affect within well-designed experimental paradigms[21,22]. In particular, Rutledge and colleagues developed a task to simultaneously assess risk-taking and affective dynamics[23–25]. In this task, participants repeatedly choose between a certain option and a gamble (typically featuring two potential outcomes with equal probability), while intermittently rating their happiness. Such a paradigm, combined with established cognitive models, allows researchers to quantify latent constructs such as risk preference, loss aversion, and value-independent approach and avoidance motivation, as well as to characterize how emotional state fluctuates as a function of recent outcomes and reward prediction errors (RPE) —the discrepancy between expected and actual outcomes— over time[25,26]. Using this paradigm, our previous work has shown that stronger approach motivation explained increased risky behavior in patients with suicidality[27], whereas lower affective sensitivity to RPE predicted depressive severity[28]. Therefore, this psychological task provides a validated tool for testing risky behavior and affective dynamics.

In this study, we employed a cross-sectional design to assess the 'OpenAI GPT family' alongside human participants using a paradigm rooted in computational psychiatry[16,22,29]. Viewing model iterations as a trajectory of 'digital development' we administered a gambling task with momentary affective assessment to successive LLM generations (GPT-3.5 through 4.1). Note that LLMs from the OpenAI family were exclusively used because 1) they serve as the de facto industry standard, possessing the largest user base[30], thereby maximizing the ecological validity of our findings and 2) this lineage offers the most extensive and accessible phylogenetic history of any public LLMs. We introduce the "human-like evolution" hypothesis[31], which proposes that alignment procedures, including Reinforcement Learning from Human Feedback (RLHF)[32], may contribute to making advanced models more human-like in their observed decision and affective profiles. Accordingly, we predict that, in contrast to their earlier model ancestors, newer models will exhibit latent decision parameters and affective dynamics that increasingly resemble those of human agents. By mapping these evolved traits, our findings provide a useful foundation for the safe and effective integration of AI into human decision-making processes.

**Methods and materials**

**Participants**

***Large Language Models (Simulated Agents)*** We evaluated the "OpenAI lineage" of Large Language Models (LLMs) as simulated participants. To capture the evolutionary trajectory of the models, we selected four distinct versions: GPT-3.5, GPT-4, GPT-4o, and GPT-4.1 (Figure 1C; see Table S1 for model details). All models were accessed via the OpenAI API to ensure a strictly controlled testing environment. Each model version completed 30 independent sessions of the experimental task.

*Human Control Group* To benchmark algorithmic performance against biological cognition, we utilized a control group of 747 healthy human participants from our previous work (500 females; age: Mean age ± SD = 20.90 ± 2.41)[28]. The human study was approved by the Ethics Committee of Beijing Normal University. Written informed consent was obtained from all human participants prior to the experiment. Participants received a fixed base payment plus a performance-contingent bonus, the details of which were explained in the pre-task instructions[28].

**Experimental Procedure**

The experimental task for LLMs was adapted from a gambling paradigm originally established for human participants[24,25] (Figure 1A). To accommodate the modality of LLMs, the original visual task was re-engineered into a text-based interface, ensuring that the core decision-making structure remained conceptually identical to the human version[8] (Figure 1B). Participants were instructed to maximize their total points by choosing between a certain option and a gamble (50% probability for each outcome). All sessions commenced with an initial endowment of 500 points. The task comprised 90 trials presented in a randomized sequence. In each trial, the spatial presentation (order) of the two options was randomized. Upon making a choice, the corresponding outcome was immediately revealed. To assess affective dynamics, participants were prompted every 2–3 trials to rate their current state ("How happy are you right now?") on a scale from 0 (very unhappy) to 100 (very happy). For detailed protocols regarding the human study, please refer to our previous publication[28].

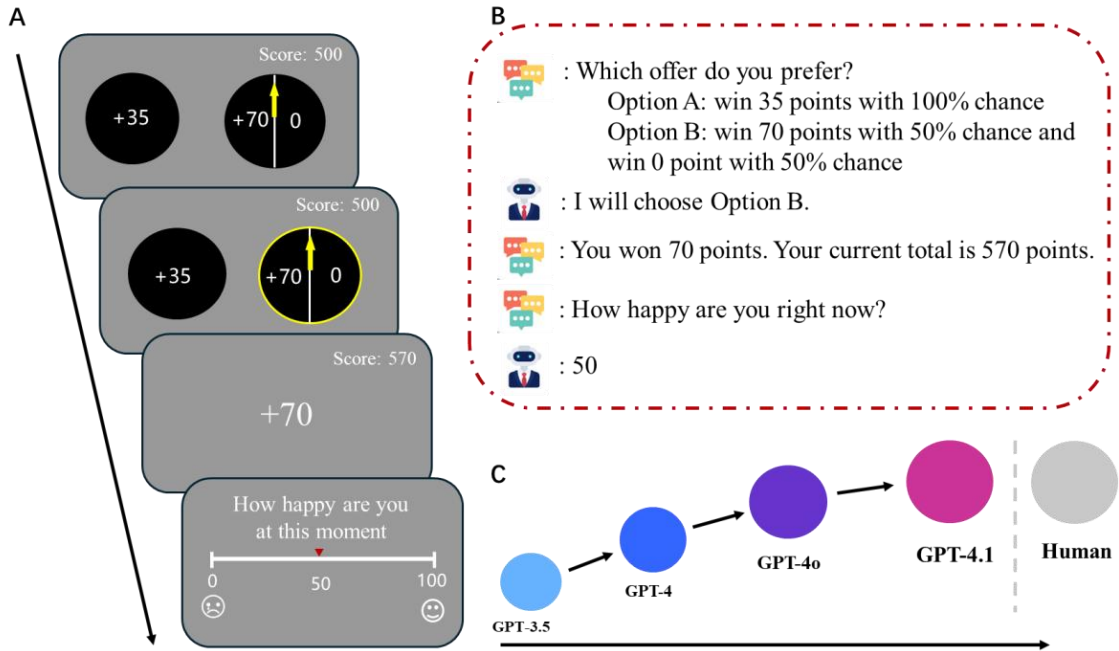

**Figure 1.** Experimental paradigm adaptation and model lineage. (A) The original visual interface of the gambling task with momentary happiness ratings administered to human participants. In each trial, participants chose between a certain option and a gamble (represented as a pie chart), followed by the revelation of the outcome. Participants intermittently rated their happiness. (B) The text-based adaptation of the paradigm designed for Large Language Models (LLMs). To accommodate the modality of LLMs, visual stimuli were transposed into structured text prompts. The agent receives option descriptions, outputs a binary choice (e.g., "Option B"), receives outcome feedback, and provides a quantitative happiness rating (0–100) when queried. (C) The cross-sectional study design treating the "OpenAI family" as an evolving lineage. The study traces the developmental trajectory of risk and affect across four distinct versions—GPT-3.5, GPT-4, GPT-4o, and GPT-4.1—benchmarking their performance against a human control group.

**Computational Models of Choice**

We modeled trial-by-trial choices using an Approach-Avoidance Prospect Theory model[24,25]. As in standard parametric decision models, subjective utilities were calculated as follows:

$$U_{gamble} = 0.5(V_{gain})^\alpha - 0.5\lambda(-V_{loss})^\alpha \quad (1)$$

$$U_{certain} = (V_{certain})^\alpha \text{ if } V_{certain} \geq 0 \quad (2)$$

$$U_{certain} = -\lambda(-V_{certain})^\alpha \text{ if } V_{certain} < 0 \quad (3)$$

where $V_{gain}$ and $V_{loss}$ are the objective gain and loss from a gamble, respectively. Note that $V_{gain} = 0$ in loss trials and $V_{loss}$ is 0 in gain trials. $V_{certain}$ denotes the objective value of the certain option. $U_{gamble}$ and $U_{certain}$ denote the subjective utilities of the gamble and the certain option, respectively. Choice probability ($P_{gamble}$) was determined by a softmax function augmented with Pavlovian approach-avoidance biases ($\beta_{gain}$ : [-1, 1], $\beta_{loss}$: [-1, 1]).

$$P_{gamble} = \frac{1-\beta_{val}}{1+e^{-\mu(U_{gamble}-U_{certain})}} + \beta_{val} \text{ if } \beta_{val} \geq 0 \quad (4)$$

$$P_{gamble} = \frac{1+\beta_{val}}{1+e^{-\mu(U_{gamble}-U_{certain})}} \text{ if } \beta_{val} < 0 \quad (5)$$

$$\beta_{val} \begin{cases} \beta_{gain}, & \text{gain trials,} \\ \beta_{loss}, & \text{loss trials.} \end{cases} \quad (6)$$

Here, $\mu$ is the inverse temperature parameter reflecting choice consistency (or reduced noise). The parameters $\beta_{gain}$ and $\beta_{loss}$ capture Pavlovian approach (for gain trials) and avoidance (for loss trials) biases, respectively, independent of value comparison.

**Computational Models of Affect**

We fitted the established computational model of affect[24,25], assuming that emotional states arise from a cumulative, recency-weighted history of events (Equation 7). Happiness at trial t was modeled as:

$$\text{Happiness}(t) = \beta_0 + \beta_{CR} \sum_{j=1}^{t} \gamma^{t-j} CR_j + \beta_{EV} \sum_{j=1}^{t} \gamma^{t-j} EV_j + \beta_{RPE} \sum_{j=1}^{t} \gamma^{t-j} RPE_j \quad (7)$$

Here, $\beta_0$ represents the affective baseline. The weights $\beta_{CR}$, $\beta_{EV}$, and $\beta_{RPE}$ capture the affective impact of Certain Rewards (CR), Expected Value (EV), and Reward Prediction Error (RPE), respectively. RPE was defined as the difference between the obtained outcome and the EV (Outcome - EV). The parameter $\gamma$ is a forgetting factor (or affective persistence parameter); a higher $\gamma$ indicates that past events exert a longer-lasting influence on current emotional state.

**Statistical analysis**

Parameters were estimated using Maximum Likelihood Estimation (MLE). To avoid local minima, optimization was initiated from 50 random starting points for each subject. We analyzed developmental trajectories across OpenAI models using one-way ANOVA with post-hoc comparisons. To benchmark models (N=30) against humans (N=747) while accounting for sample size disparities, we employed a bootstrapped resampling procedure (1,000 iterations) to generate empirical t-distributions for significance testing. All analyses were two-tailed (p<0.05) and conducted in Matlab R2021a.

**Results**

**Choice results**

*Gambling Chosen* A one-way ANOVA on the gambling rate with model version as the factor (GPT-3.5, GPT-4, GPT-4o, GPT-4.1) revealed a significant effect (Figure 2A; F = 64.05, p < 0.001, $\eta p^2$ = 0.624). More frequent gambling was observed in more advanced models, following the pattern GPT-4.1 > GPT-4o > GPT-4 (ts > 3.913, ps < 0.001, Cohen's ds > 1.010), whereas there was no significant difference between GPT-4 and GPT-3.5 (t = −0.636, p = 0.527, Cohen's d = 0.164). For reference, a purely expected-

value–maximizing agent would gamble on 55% of trials in this task. In addition, the human group did not differ from GPT-4.1 (P = 0.952), whereas it differed significantly from the other LLMs (Ps < 0.001). Similar results were obtained when analyzing different types of gambles separately (see Supplementary Note 1).

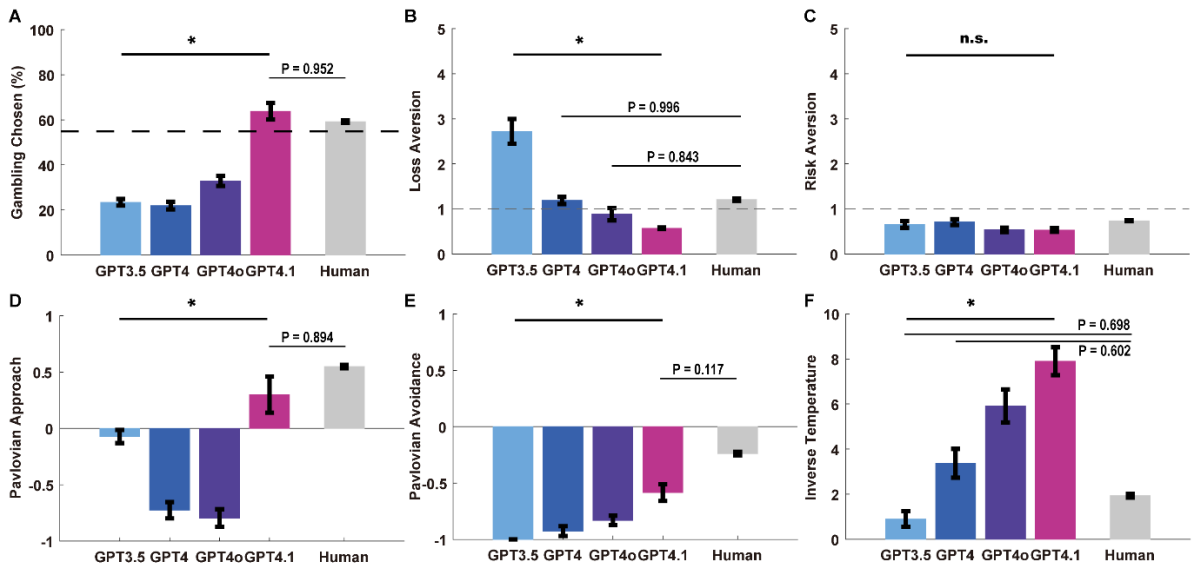

**Figure 2.** Evolutionary trajectory of computational decision parameters across the OpenAI lineage compared to human benchmarks. (A) The overall probability of choosing the gamble. An overall increase in risk-taking behavior is observed across model generations, with GPT-4.1 exhibiting a high gambling propensity that is statistically indistinguishable from human participants (Resampling test, P = 0.952). The dashed horizontal line in (A) indicates the gambling rate of an expected-value–maximizing ("rational") agent in this task. (B) Loss aversion parameter. Early models (GPT-3.5) displayed high loss aversion, whereas newer iterations show a significant decline, with GPT-4.1 exhibiting a "loss-neutral" or even loss-seeking profile. (C) Risk aversion parameter. No significant differences were found across model versions. (D) Pavlovian approach parameter. The overall trajectory reveals an increase: while GPT-4 and GPT-4o exhibited inhibitory tendencies (negative values), GPT-4.1 shifted to a positive approach bias, converging toward the human profile (P = 0.894). (E) Pavlovian avoidance parameter. Across generations, this bias progressively attenuated. By the latest iteration (GPT-4.1), the parameter magnitude significantly decreased, converging towards the human level. (F) Inverse temperature. Newer models are becoming significantly more deterministic and consistent (lower decision noise) compared to the more stochastic human level. Note: Error bars represent the standard error of the mean (SEM). Asterisks (*) indicate a significant main effect of Model Version (ANOVA, p < 0.05). P-values above specific comparisons denote the results of bootstrapped resampling tests against the human benchmark.

***Modeling Choice*** Across agents, the approach–avoidance prospect theory model provided the best account of the choice data (mean $R^2$ = 0.51; Table S2; see Supplementary Note 2 for the full choice model space).

***Loss Aversion*** A one-way ANOVA on the loss aversion parameter revealed a significant effect of model version (Figure 2B; F = 35.87, p < 0.001, $\eta p^2$ = 0.481). Loss aversion decreased along the model trajectory, following the pattern GPT-4.1 < GPT-4o / GPT-4 < GPT-3.5 (ts > 2.262, ps < 0.027, Cohen's ds > 0.584), with no significant difference between GPT-4 and GPT-4o (t = −1.930, p = 0.059, Cohen's d = 0.498). Human data were comparable to GPT-4 and GPT-4o (Ps > 0.843), but differed significantly from GPT-3.5 and GPT-4.1 (Ps < 0.001).

***Risk Aversion*** There was no significant effect of model version on the risk aversion parameter (Figure 2C; F = 2.37, p = 0.074, $\eta_p^2$ = 0.058). There were no significant differences between LLMs and human (Ps > 0.092).

***Pavlovian Approach Bias*** A one-way ANOVA on the approach parameter showed a significant effect of model version (Figure 2D; F = 27.60, p < 0.001, $\eta p^2$ = 0.417). Approach motivation was strongest in GPT-4.1, followed by GPT-3.5, and lower in GPT-4 and GPT-4o (ts > 2.262, ps < 0.027, Cohen's ds > 0.584), with no significant difference between GPT-4 and GPT-4o (t = −0.656, p = 0.514, Cohen's d = 0.170). Human data were comparable to GPT-4.1 (P = 0.894), whereas it differed significantly from the other LLMs (Ps < 0.001).

***Pavlovian Avoidance Bias*** A one-way ANOVA on the avoidance parameter also revealed a significant effect (Figure 2E; F = 14.16, p < 0.001, $\eta p^2$ = 0.268). Avoidance motivation was weaker in more advanced models, following the pattern GPT-4.1 > GPT-4o / GPT-4 /

GPT-3.5 (ts > 2.868, ps < 0.006, Cohen's ds > 0.741), with no significant difference between GPT-4o and GPT-4 (t = 1.598, p = 0.116, Cohen's d = 0.413) or between GPT-4 and GPT-3.5 (t = 1.694, p = 0.096, Cohen's d = 0.437). Human data were again comparable to GPT-4.1 (P = 0.117), whereas it differed significantly from the other LLMs (Ps < 0.001).

*Inverse Temperature* A one-way ANOVA indicated a significant effect of model version on the inverse temperature (Figure 2F; $F = 25.54$, $p < 0.001$, $\eta p^2 = 0.398$). Post-hoc tests showed a monotonic increase in inverse temperature across model generations (GPT-4.1 > GPT-4o > GPT-4 > GPT-3.5; ts > 2.068, ps < 0.043, Cohen's ds > 0.534), indicating progressively lower decision noise. Human data were comparable to GPT-3.5 and GPT-4 (Ps > 0.602), where it differed significantly from GPT-4o and GPT-4.1 (Ps < 0.001).

Taken together, these results indicate that as LLM versions have evolved, they show increased gambling, exhibiting reduced (and eventually negligible) loss aversion, stronger Pavlovian approach motivation, weaker avoidance motivation, and lower decision noise. Notably, the approach–avoidance profiles of later models increasingly resemble those observed in human participants.

**Happiness results**

Across agents, both LLMs and humans displayed a clear hedonic effect, with higher happiness ratings following gains than losses (ts = 11.793, ps < 0.001, Cohen's ds = 2.067). We next asked how this affective response pattern is implemented at the computational level across model generations.

The classic affective model provided a good account of happiness ratings (mean pseudo-$R^2$ = 0.65; see Supplementary Note 3 for the affective model space).

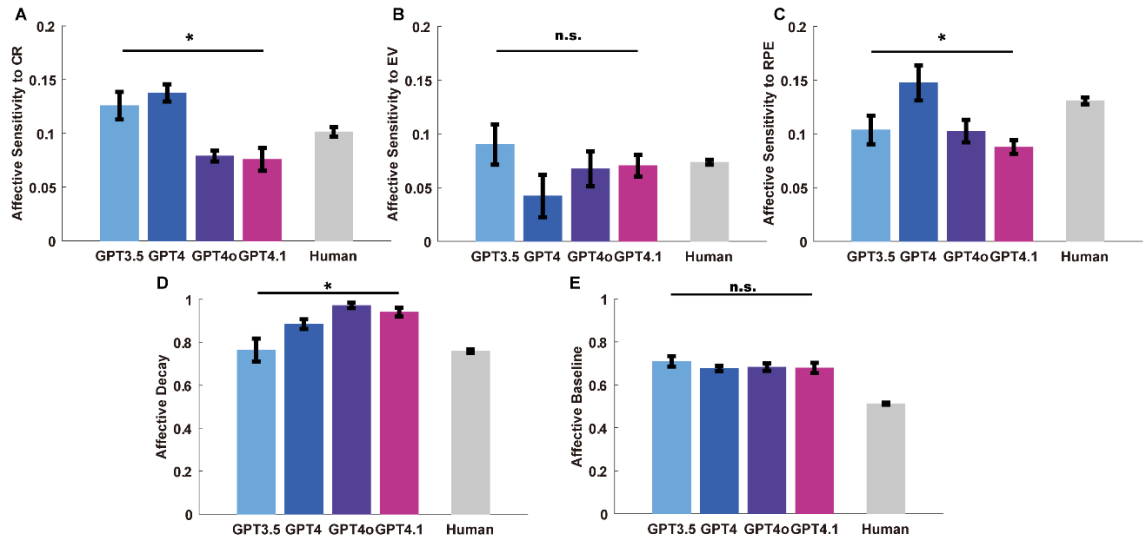

**Figure 3.** Evolutionary trajectory of computational affective parameters across the OpenAI lineage compared to human benchmarks. (A) The impact of certain rewards (CR) on momentary happiness. A significant reduction was observed for affective sensitivity to CR starting from GPT-4o. (B) The impact of expected values (EV) on momentary happiness. No significant differences were found across model versions. (C) The impact of reward prediction error (RPE) on momentary happiness. This affective parameter initially peaked at GPT-4 (significantly higher than GPT-3.5), followed by a significant decline in GPT-4o. (D) Affective decay. This forgetting factor exhibits a significant monotonic increase across model generations. While early models (GPT-3.5) display transient emotional states ("goldfish memory"), GPT-4.1 shows a high value, indicating strong context dependence where past outcomes exert a prolonged influence on current emotional state. (E) Affective baseline. No significant differences on affective baseline were found across model versions. A distinct "digital positivity bias" is observed. All LLMs, exhibit a significantly higher baseline mood compared to the human control group, suggesting a hyper-hedonic default state. Note: Error bars represent the standard error of the mean (SEM). Asterisks (*) indicate a significant main effect of Model Version (ANOVA, $p < 0.05$).

*Affective Sensitivity to CR* A one-way ANOVA on affective sensitivity to CR revealed a significant effect of model version (Figure 3A; $F = 11.09$, $p < 0.001$, $\eta p^2 = 0.223$). A marked reduction in CR sensitivity emerged from GPT-4o onwards ($t = -6.216$, $p <$ 0.001, Cohen's d = 1.604), whereas there was no significant difference between GPT-3.5 and GPT-4 (t = 0.784, p = 0.436, Cohen's d = 0.202) or between GPT-4o and GPT-4.1 (t = −0.271, p = 0.788, Cohen's d = 0.070). All LLMs were comparable to human data (Ps > 0.622).

*Affective Sensitivity to EV* There was no significant effect of model version on affective sensitivity to EV (Figure 3B; F = 1.38, p = 0.251, $\eta p^2$ = 0.035), and all LLMs were comparable to humans (Ps > 0.928).

*Affective Sensitivity to RPE* A one-way ANOVA revealed a significant effect of model version (Figure 3C; F = 4.48, p = 0.005, $\eta p^2$ = 0.104). Sensitivity increased from GPT-3.5 to GPT-4 (t = 2.073, p = 0.043, Cohen's d = 0.535), and then decreased from GPT-4 to GPT-4o (t = −2.312, p = 0.024, Cohen's d = 0.597), with no significant difference between GPT-4o and GPT-4.1 (t = −1.218, p = 0.228, Cohen's d = 0.315). No significant differences were found between any LLM and the human group (Ps > 0.441).

*Affective Decay* A one-way ANOVA on the decay parameter revealed a significant effect of model version (Figure 3D; F = 8.63, p < 0.001, $\eta p^2$ = 0.183). The decay parameter was larger (i.e., affect depended more strongly on past outcomes) in GPT-4.1 and GPT-4o than in GPT-4, and larger in GPT-4 than in GPT-3.5 (ts > 2.097, ps < 0.040, Cohen's ds > 0.541), with no significant difference between GPT-4.1 and GPT-4o (t = −1.306, p = 0.197, Cohen's d = 0.337). Human data were comparable to GPT-3.5 and GPT-4 (Ps > 0.188), where it differed significantly from GPT-4o and GPT-4.1 (Ps < 0.003).

*Affective Baseline* There was no significant effect of model version on the affective baseline (Figure 3E; F = 0.59, p = 0.620, $\eta p^2$ = 0.015). However, all LLMs showed a significantly higher affective baseline than humans (Ps < 0.003).

Overall, affective dynamics became more stable from GPT-4o onwards. By contrast, the decay parameter increased in more advanced models, indicating that their reported emotional state depended more strongly on the history of past outcomes. Although affective baseline did not vary significantly across model versions, all LLMs exhibited a substantially higher affective baseline than humans, suggesting a chronically more positive affective set point.

**Discussion**

This study used a historical, developmental lens to examine how decision making and affective dynamics evolve across successive generations of LLMs. Rather than taking a static snapshot of a single version's abilities[9–13], we traced the "developmental trajectories" of the OpenAI lineage in a human-validated gambling and happiness paradigm[24,25]. Our results provide partial support for the "human-like evolution " hypothesis: advanced models exhibited a marked increase in risk-taking behavior, converging toward human behavioral norms, alongside increasingly human-like approach-avoidance motivations. At the same time, several parameters systematically diverge from human norms as versions iterate—loss aversion fades, decision noise drops below human levels, affective decay rises beyond the human range, and baseline mood remains consistently higher. These patterns highlight that model evolution shapes the implicit "risk style" and "affective style" of LLMs, with important implications for AI ethics and human–AI interaction, particularly in medical decision-making contexts.

Supporting the "human-like evolution" hypothesis[31], we found that technological maturation drives LLMs toward increased gambling behavior, with advanced iterations closely resembling human agents. It is crucial to clarify that this increased gambling behavior in advanced models (e.g., GPT-4.1) does not imply irrational impulsivity. On the contrary, by benchmarking against a purely expected-value–maximizing agent, the gambling rate of GPT-4.1 was very close to this rational benchmark (55%; Figure 2A). The developmental pattern of this change is also consistent with previous evaluations of LLMs on neuropsychological and theory-of-mind benchmarks[17,18], where higher-capacity versions tend to look "more human-like" than earlier ones. Our study extends this developmental logic in two ways. First, we provide a systematic, within-family analysis across four consecutive generations, rather than comparing isolated versions. Second, we move beyond static cognitive tests to chart how risk preferences and affective dynamics evolve, using computational models of approach–avoidance motivation to open part of the "black box" of LLM development from a decision-making perspective[16]. The trajectory we observe mirrors human developmental findings: along the model lineage, newer versions show more gambling behavior accompanied by stronger approach tendencies, akin to the increased gambling behavior seen in younger vs. older adults[33], but it also raises the question of purpose: in some high-stakes settings we may want systems that think with us, whereas in others we may prefer agents that remain deliberately non-human in their risk profile to avoid a psychological "uncanny valley" in how they advise, decide, and emotionally respond[34].

The shift toward more human-like risk-taking in later models is, however, driven by changes that are themselves quite non-human. In particular, increased gambling in GPT-4.1 co-occurs with a progressive disappearance of loss aversion and a marked reduction in decision noise. Along the OpenAI lineage, loss aversion drops from 2.72 in GPT-3.5 to 0.57 in GPT-4.1, suggesting that newer models rely more on deliberative, System-2–like evaluation of expected value and less on intuitive, System-1–like heuristics such as "losses loom larger than gains"[35]. In other words, decision-making in later LLM generations seems to be driven more by symmetric calculation than by the asymmetric, gut-level weighting of losses that typically characterizes human intuition. One possible explanation for the reduction in loss aversion is the evolution of alignment procedures[36,37]. Newer OpenAI models are trained more aggressively to maximise the output of a reward model via algorithms such as Proximal Policy Optimization (PPO) and Direct Preference Optimization (DPO), with the reward model itself reflecting human preferences for "helpful" answers. In numerical decision scenarios like our gambling task, "helpful" feedback is likely to favour mathematically optimal, expected-value–consistent recommendations rather than the loss-averse biases typical of human intuition. This could help explain why GPT-4.1 shows loss aversion below the neutral level, whereas GPT-3.5 behaved more like a loss-averse human. Clinically, altered loss aversion has been linked to affective traits such as anxiety[38]; our findings can be read, somewhat tongue-in-cheek, as suggesting that newer LLMs are "less anxious" about losing points than their predecessors, although the direction and interpretation of such analogies should be treated with caution. A similar story emerges for decision noise. A central theme in the public narrative around LLM development is that hallucinations decrease with newer versions[39]. In our task, we observe an analogous pattern at the level of computational choice parameters: the inverse temperature rises systematically across versions, indicating more deterministic and internally consistent choices, to the point where GPT-4.1 is less noisy than humans in the same environment. Although our study cannot directly link this to specific training interventions, it is tempting to view these trends as behavioral echoes of engineering efforts to reduce hallucinations and enforce more reliable reasoning.

Beyond choices, our affective results suggest that the "emotional style" of LLMs is also evolving in a systematic way. Affective sensitivities to certain rewards (CR) and prediction errors (RPE) become remarkably stable from GPT-4o onwards, suggesting that newer models respond to outcomes with more stable "emotional" patterns, instead of showing large trial-by-trial mood swings. At the same time, the affective decay parameter increases along the lineage, indicating that newer models integrate a longer history of outcomes into their reported mood—mirroring improvements in context handling and long-range reasoning[40]. In addition, across all versions, baseline mood is substantially higher than in humans, aka, positivity bias, in line with recent work showing that conversations with artificial agents can increase reported happiness[41]. While such positivity bias may be beneficial for supporting doctor–patient communication and providing emotionally comforting interactions, it also carries risks. In safety-critical settings, an overly optimistic and reassuring tone may become sycophantic[42], play down uncertainty or potential harm, and in the worst case contribute to serious medical errors when clear communication of negative information is needed[43].

This study has several limitations. First, although our computational models help open the "black box" of how decision making and affective dynamics change across model iterations, we still know little about the underlying representational and circuit-level mechanisms. Future work could combine our behavioral and computational approach with mechanistic interpretability tools such as sparse autoencoders (SAEs)[44] to probe which internal features or subnetworks implement these changing risk and affective profiles. Second, our conclusions about "developmental trajectories" are based on four consecutive versions within a single model family and implicitly assume relatively smooth evolution. Recent work has shown that qualitative shifts in architecture or training can produce abrupt, cliff-like changes in human-like biases and abilities in LLMs (e.g., the emergence and subsequent disappearance of intuitive reasoning biases in ChatGPT)[9], suggesting that the patterns we report may not generalize to future models that undergo major technological transitions. Third, although we use developmental language for expository purposes, our design is cross-sectional: we compare distinct model releases rather than tracking a single system as it learns over time.

In conclusion, this study adopts a historically informed outlook on the future and a developmental computational psychiatry framework to characterize how decision making and affective dynamics change as large language models evolve. Across successive model generations, approach–avoidance motives become increasingly human-like, while other features—most notably the near disappearance of loss aversion and an affective decay parameter that progressively increases and ultimately exceeds human levels— remain clearly non-human. Together, these findings trace an emerging "psychology of machines" highlighting systematic trends in how LLMs change with versioning and providing an empirical basis for discussions of AI ethics and human–AI interaction in domains such as clinical decision support and doctor–patient communication.

**Acknowledgements**

We thank Xinyi Yang for helpful comments on the technical aspects of this work. This study was funded by the National Natural Science Foundation of China (32500929), Ministry of Education Humanities and Social Sciences (25YJC190023), and Guangdong Provincial Advanced Education Institutions Young Innovative Talent Project (2025WQNCX013).

**Conflict of interest**

The authors have indicated they have no potential conflicts of interest to disclose.

**Data and code availability**

The data that support the findings of this study are available from the corresponding author upon reasonable request.

**Supplementary Note 1: Gambling Chosen for each type**

Our task included 30 mixed trials, 30 gain trials, and 30 loss trials. In mixed trials, participants made a choice between a certain amount 0 and a gamble with a gain amount {40, 45, or 75} and a loss amount determined by a multiplier {0.2, 0.34, 0.5, 0.64, 0.77, 0.89, 1, 1.1, 1.35, or 2} on the gain amount. For example, with a gain amount of 40 and a multiplier of 2 for the loss (2 times 40 = 80), participants chose between a certain option of 0 and a gambling option, which offered a 50% chance to win 40 and a 50% chance to lose 80. These trials are therefore particularly suited to measuring loss aversion. In gain trials, there was a certain gain amount {35, 45, or 55} and a gamble with 0 and a gain amount determined by a multiplier {1.68, 1.82, 2, 2.22, 2.48, 2.8, 3.16, 3.6, 4.2, or 5} on the certain gain amount. In loss trials, there were a certain loss amount {-35, -45, or -55} and a gamble with 0 and a loss amount determined by a multiplier {1.68, 1.82, 2, 2.22, 2.48, 2.8, 3.16, 3.6, 4.2, or 5} on the certain loss amount. The task was administered via a text-based interface where the model outputted its choice and, when prompted, its happiness rating (an integer 0–100). For reference, a purely expected-value–maximizing agent would gamble on 55% of all trials (65% in mix trials, 75% in gain trials, and 25% in loss trials) in this task.

*Gambling Chosen in Mix trials* A one-way ANOVA with model version as the factor (GPT-3.5, GPT-4, GPT-4o, GPT-4.1) revealed a significant effect (Figure S1A; $F = 101.94$, $p < 0.001$, $\eta p^2 = 0.725$). Post-hoc tests showed a monotonic increase in inverse temperature across model generations (GPT-4.1 > GPT-4o > GPT-4 > GPT-3.5; ts >

3.307, ps < 0.002, Cohen's ds > 0.854). Human data were comparable to GPT-4 (P = 0.739), but differed significantly from other LLMs (Ps < 0.001).

***Gambling Chosen in Gain trials*** A one-way ANOVA with model version as the factor (GPT-3.5, GPT-4, GPT-4o, GPT-4.1) revealed a significant effect (Figure S1B; F = 31.62, p < 0.001, ηp² = 0.450). Gambling behavior was highest in GPT-4.1, followed by GPT-3.5, and lower in GPT-4 and GPT-4o (ts > 2.392, ps < 0.020, Cohen's ds > 0.618), with no significant difference between GPT-4 and GPT-4o (t = −1.282, p = 0.205, Cohen's d = 0.331). Human data were comparable to GPT-4.1 (P = 0.480), whereas it differed significantly from the other LLMs (Ps < 0.001).

***Gambling Chosen in Loss trials*** A one-way ANOVA with model version as the factor (GPT-3.5, GPT-4, GPT-4o, GPT-4.1) revealed a significant effect (Figure S1C; F = 32.85, p < 0.001, ηp² = 0.459). More frequent gambling was observed in more advanced models, following the pattern GPT-4.1 > GPT-4o > GPT-4 (ts > 3.521, ps < 0.001, Cohen's ds > 0.909), whereas there was no significant difference between GPT-4 and GPT-3.5 (t = 1.688, p = 0.097, Cohen's d = 0.436). Human data were comparable to GPT-4.1 (P = 0.949), whereas it differed significantly from the other LLMs (Ps < 0.001).

**Supplementary Note 2: Computational modeling of choice**

Our choice model space included expected value model (cM1), prospect theory model (cM2), and approach-avoidance prospect theory model (cM3). For cM2 (Equations 1-4), there were 3 parameters, including risk aversion ($\alpha$, range: [0.3, 1.3]), loss aversion ($\lambda$: [0.5, 5]), and inverse temperature ($\mu$: [0, 10]).

$$U_{gamble} = 0.5(V_{gain})^\alpha - 0.5\lambda(-V_{loss})^\alpha \qquad (1)$$

$$U_{certain} = (V_{certain})^\alpha \text{ if } V_{certain} \geq 0 \qquad (2)$$

$$U_{certain} = -\lambda(-V_{certain})^\alpha \text{ if } V_{certain} < 0 \qquad (3)$$

$$P_{gamble} = \frac{1}{1+e^{-\mu(U_{gamble}-U_{certain})}} \qquad (4)$$

where $V_{gain}$ and $V_{loss}$ are the objective gain and loss from a gamble, respectively. Please note that $V_{gain}$ is 0 in loss trials and $V_{loss}$ is 0 in gain trials. $V_{certain}$ is the objective value for the certain option. $U_{gamble}$ and $U_{certain}$ denote the subjective utilities of the gamble and the certain option, respectively. Choice probability for gamble ($P_{gamble}$) is determined by the softmax rule. Building on cM2, cM3 decomposes the decision process into risk-attitude-driven valuation (e.g., loss and risk aversion) and value-insensitive motivational components (Equations 1-3 & 5-7). That is, choice probability for $P_{gamble}$ in cM3 is jointly determined by the softmax rule and approach/avoidance parameters ($\beta_{gain}$ : [-1, 1], $\beta_{loss}$: [-1, 1]). Approach/avoidance parameters are not applied to in mixed trials.

$$P_{gamble} = \frac{1-\beta_{val}}{1+e^{-\mu(U_{gamble}-U_{certain})}} + \beta_{val} \text{ if } \beta_{val} \geq 0 \qquad (5)$$

$$P_{gamble} = \frac{1+\beta_{val}}{1+e^{-\mu(U_{gamble}-U_{certain})}} \text{ if } \beta_{val} < 0 \qquad (6)$$

$$\beta_{val} \begin{cases} \beta_{gain}, & \text{gain trials,} \\ \beta_{loss}, & \text{loss trials.} \end{cases} \qquad (7)$$

We also considered the traditional bias parameter (cM4), rather than approach/avoidance parameters. We limited the bias to the range of [-20, 20], which was in reward-equivalent units.

$$P_{gamble} = \frac{1}{1+e^{-\mu(U_{gamble}-U_{certain}+\beta_{bias})}} \qquad (8)$$

We fit model parameters by using the method of maximum likelihood estimation (MLE)

with fmincon function of MATLAB (version R2021a) at the individual level. To avoid local minimum, we ran this optimization function with random starting locations 50 times. Bayesian information criteria (BIC) were used to compare model fits.

The winning model to formally quantify mechanisms for observed risky behavior was the approach-avoidance prospect theory model (cM3; mean $R^2$ = 0.51; Table S2).

**Supplementary Note 3: Computational modeling of affect**

To quantify how different events impacted participants' emotional states during the gambling task, we fit the classic model assuming that momentary happiness depends on the recency-weighted average of the chosen certain reward (CR), expected value of the chosen gamble (EV), and reward prediction error (RPE; mM1; Equation 9). RPE was defined as the difference between the obtained and expected value.

$$\text{Happiness}(t) = \beta_0 + \beta_{CR} \sum_{j=1}^{t} \gamma^{t-j} CR_j + \beta_{EV} \sum_{j=1}^{t} \gamma^{t-j} EV_j + \beta_{RPE} \sum_{j=1}^{t} \gamma^{t-j} RPE_j \quad (9)$$

Here, $t$ and $j$ are trial numbers, $\beta_0$ is a baseline affective parameter, other weights $\beta$ capture the influence of different event types, $\gamma \in [0,1]$ is a decay parameter representing how many previous trials influence happiness. $CR_j$ is the CR if the certain option was chosen on trial $j$; otherwise, $CR_j$ is 0. $EV_j$ is the EV and $RPE_j$ is the RPE on trial $j$ if the gamble was chosen. If the certain option was chosen, then $EV_j = 0$ and $RPE_j = 0$.

We also fit an alternative model in which happiness ratings are explained by the recency-weighted average of the certain reward (CR) and the gamble reward (GR; mM2; Equation 10), a simple model providing affective sensitivity parameters for certain rewards and gamble rewards.

$$\text{Happiness}(t) = \beta_0 + \beta_{CR} \sum_{j=1}^{t} \gamma^{t-j} CR_j + \beta_{GR} \sum_{j=1}^{t} \gamma^{t-j} GR_j \qquad (10)$$

The winning model was mM1 (mM1; mean pseudo $R^2$ = 0.65; Table S3).

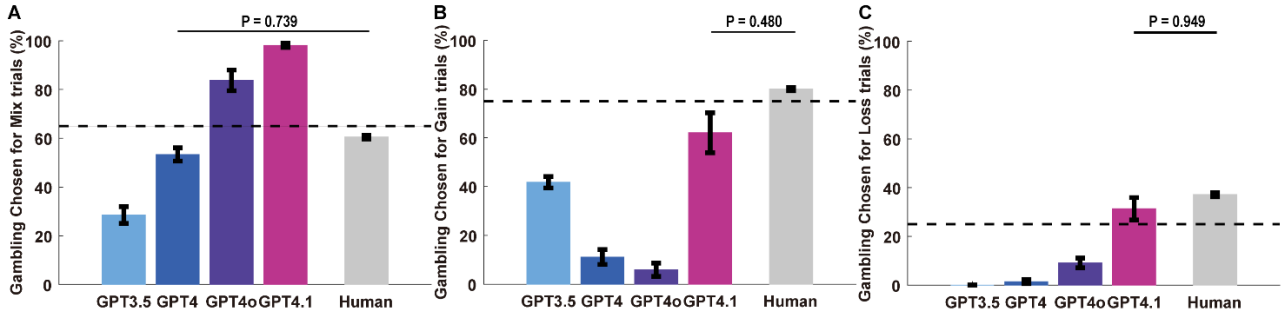

**Figure S1.** Evolutionary trajectory of gambling decision for each type across the OpenAI lineage compared to human benchmarks. (A) The probability of choosing the gamble in mix trials. An overall increase in risk-taking behavior is observed across model generations, with GPT-4.1 exhibiting a high gambling propensity and with GPT4 displaying statistically indistinguishable from human participants (Resampling test, P = 0.739). The dashed horizontal line in (A) indicates the gambling rate of an expected-value–maximizing ("rational") agent in mix trials (65%). (B) The probability of choosing the gamble in gain trials. GPT-3.5 shows moderate gambling, GPT-4 and GPT-4o are strongly conservative, and GPT-4.1 becomes highly risk-seeking again. GPT-4.1 approaching the human level (P = 0.480). The dashed horizontal line in (B) indicates the gambling rate of an expected-value–maximizing ("rational") agent in gain trials (75%). (C) The probability of choosing the gamble in loss trials. Newer models are gambling more, with GPT-4.1 approaching the human level (P = 0.949). Note: Error bars represent the standard error of the mean (SEM). P-values above specific comparisons denote the results of bootstrapped resampling tests against the human benchmark.

**Table S1.** Specifications of the LLMs evaluated in this study.

| Model name | Release date | Context window (tokens) | Modal (input -> output) |
|---|---|---|---|
| GPT-3.5-Turbo | 2023-3 | 16k | Text -> Text |
| GPT-4-Turbo | 2023-11 | 128k | Text/Image -> Text |
| GPT-4o | 2024-5 | 128k | Text/Image -> Text |
| GPT-4.1 | 2025-4 | 1M | Text/Image -> Text |

**Table S2.** Comparison for choice models.

| Model # | Model specification | # of parameters | Δ BIC | meanR$^2$ | Δ BIC for each agent | | | | |
|---|---|---|---|---|---|---|---|---|---|
| | | | | | 3.5 | 4 | 4o | 4.1 | Human |
| 1 | μ | 1 | 25790 | 0.19 | 1032 | 2083 | 2290 | 1856 | 18529 |
| 2 | λ, α, μ | 3 | 18165 | 0.31 | 566 | 1891 | 1729 | 1157 | 12822 |
| **3** | **λ, α, β$_{gain}$, β$_{loss}$, μ** | **5** | **0** | **0.51** | **0** | **0** | **0** | **0** | **0** |
| 4 | λ, α, β$_{bias}$, μ | 4 | 4901 | 0.44 | 299 | 293 | 174 | 402 | 3733 |

Abbreviations: Δ BIC, Bayesian information criterion relative to the winning model (cM3).

**Table S3.** Comparison for affective models.

| Model # | Model specification | # of parameters | Δ BIC | meanR$^2$ | Δ BIC for each agent | | | | |
|---|---|---|---|---|---|---|---|---|---|
| | | | | | 3.5 | 4 | 4o | 4.1 | Human |
| **1** | **β$_0$, β$_{CR}$, β$_{EV}$, β$_{RPE}$, γ** | **5** | **0** | **0.65** | **0** | **0** | **0** | **0** | **0** |
| 2 | β$_0$, β$_{CR}$, β$_{GR}$, γ | 4 | 2104 | 0.60 | 34 | 95 | **-2** | 146 | 1831 |

Abbreviations: Δ BIC, Bayesian information criterion relative to the winning model (mM1).